\documentclass[aps,pre,twocolumn,groupedaddress]{revtex4-1}

  \usepackage{graphicx}

\begin{document}

\title{Blocking a wave:  Frequency band gaps in ice shelves with periodic crevasses }

\author{Julian Freed-Brown$^1$, Jason M. Amundson$^2$, Douglas R. MacAyeal$^2$, Wendy W. Zhang$^1$}

\affiliation{
$^1$ Department of Physics and the James Franck Institute, University of Chicago, 929 E. 57th Street, Chicago, IL, 60637, USA.\\
$^2$ Department of Geophysical Sciences, University of Chicago, 5734 S. Ellis Ave, Chicago, IL, 60637, USA.}

\begin{abstract}
We assess how the propagation of high-frequency elastic-flexural waves through an ice shelf is modified by the presence of spatially periodic crevasses.   Analysis of the normal modes supported by the ice shelf with and without crevasses reveals that a periodic crevasse distribution qualitatively changes the mechanical response.  The normal modes of an ice shelf free of crevasses are evenly distributed as a function of frequency.  In contrast, the normal modes of a crevasse-ridden ice shelf are distributed unevenly. There are ``band gaps", frequency ranges over which no eigenmodes exist.   A model ice shelf that is $50$ km in lateral extent and $300$ m thick with crevasses spaced $500$ m apart has a band gap from $0.2$ to $0.38$ Hz.  This is a frequency range relevant for ocean wave/ice-shelf interactions.  When the outermost edge of the crevassed ice shelf is oscillated at a frequency within the band gap, the ice shelf responds very differently from a crevasse-free ice shelf.  The flexural motion of the crevassed ice shelf is confined to a small region near the outermost edge of the ice shelf and effectively ``blocked' from reaching the interior.
\end{abstract}

\maketitle

\section{Introduction}
Crevasses are usually thought to weaken an ice shelf. There are many works investigating the different physical mechanisms by which crevasses originate, grow, and penetrate the ice shelf \cite{VanderveenFrac}, \cite{Vanderveen98}, thereby causing calving \cite{Vanderveen02} or large-scale collapse \cite{Scambos00}.  These works typically focus on the stress distribution in the vicinity of a single crevasse.  In this work, we show that crevasses, collectively, can in fact be a source of mechanical resilience.  Specifically, we analyze the normal mode distribution and the small-amplitude, dynamical response of a model ice shelf containing a spatially periodic array of crevasses (Figure~1).  Numerical results show that the crevasse-ridden ice shelf has no normal modes from $0.2$~Hz to $0.38$~Hz, a feature known as a ``band gap" in physics and mechanics literature.  An intact ice shelf that has no crevasses would have many elastic-flexural modes over the same frequency range. The modes that reside in the band gap interval for an intact ice shelf are instead shifted to the edges of the band gap.  The net effect is that the crevassed ice shelf has a large number of excess normal modes whose natural frequencies lie near the edges of the band gap.  As a result, the response of a crevassed ice shelf has distinct peaks in the power spectrum, with peak locations corresponding to the edges of the band gap.  When the ice shelf is driven at its seaward edge at a frequency that lies within the band gap, the ice shelf motion is strongly attenuated as one moves away from the free edge. The behavior is reminiscent of an evanescent wave. Intriguingly, results from a recent study on the response of the Ross ice shelf to ocean gravity wave forcing uncovered a peak near $0.5$~Hz \cite{Bromirski11}. Thus the band gap feature studied here may be relevant for the mechanical response of real ice shelves (see also \cite{McGrath11}). 

There exists a variety of research on the propagation of classical waves through an infinite, periodic structure \cite{Sheng06}. The applications include  crystallography, phononic crystals,  and electron transport in metals and semiconductors \cite{Ashcroft}.  Within geophysical sciences, band gaps have been used to explain the formation of near-shore, underwater sandbars \cite{Mei85} and gravity wave propagation through sea ice \cite{Chou98}.

In short, band gaps are possible whenever waves propagate through a periodic structure.  To see why, consider how a transverse wave propagates along a 1D string.  This motion satisfies the linear wave equation and propagates at a uniform speed.  However, when an imperfection is present, for example a bead fixed onto the string, the wave is partially reflected and partially transmitted upon encountering the bead.  When a periodic arrangement of beads are placed along the string, the wave is reflected and transmitted partially every time it comes across a bead.   Typically, when a large number of beads are present, the phases of the reflected portions of the wave are randomly distributed and do not add together. However,  it is possible to arrange the wavelength of the transverse wave so that the phases of all the reflected waves arrive at the end of the string in phase with each other, thus adding constructively.  In this special situation,  the coherent back scattering channels almost all the energy of the incident wave into the reflected wave. In the limit of an infinite periodic array of beads along the string, the wave no longer propagates but instead becomes evanescent, i.e. decays exponentially with distance from the free edge. 

The basic ingredients for this qualitative transition are simple:  a periodic array of defects, a system whose lateral dimension is much larger than the spacing between the defects, and finally the possibility of exciting a normal mode of the system whose wavelength is comparable with the spacing between the defects.  The propagation of elastic-flexural waves along crevasse-ridden ice shelves satisfy all these criteria, and therefore supports band gaps (as we demonstrate below).  In the same spirit, we also analyze the dispersal of energy associated with seiches in a fjord when ice melange is present \cite{MacAyeal11} and show that, if the melange can be idealized by a periodic array of solid ice blocks, then a band gap structure also exists for the normal modes of the fjord. In a sense, these two cases can be viewed as different limits of the same underlying problem of how wave energy propagates through an ice-covered sea surface.

Previous works on wave/ice-shelf interactions have focused on the low-frequency, long-wavelength limit dominated by the inertia associated with water motion beneath the ice shelf.  In these works, the ice shelf is idealized as intact and simply modifies the normal stress boundary condition on the water surface (see for example \cite{Balmforth99}).  We take the opposite approach and focus on elastic-flexural waves, which are higher in frequency.  When a thin ice shelf overlies a deep layer of water, the dispersion relation for the shelf motion has the form 
	\begin{equation}
	4\pi^2 f^2 = \frac{D k^5}{\rho_w+\rho_i h k},
	\end{equation}
where $f$ is the vibrational frequency, $D=E h^3/12(1-\nu^2)$ is the flexural rigidity, $E$ is the Young's modulus, $\nu$ the Poisson ratio, $k$ is the wavenumber, $\rho_{w}$ is the density of water, and $h$ is the shelf thickness \cite{Landau}.  For the propagation of short-wavelength, elastic-flexural waves (or $h k \gg 1$), the coupling between water motion and shelf motion becomes irrelevant and the dispersion relation simplifies to
	\begin{equation}
	4\pi^2 f^2 = \frac{D k^4}{h \rho_i}. 
	\end{equation}
This is the dispersion relation of a thin elastic plate vibrating freely under a balance of elastic stresses and inertia.  Motivated by this limiting behavior, we opt to neglect the normal stress exerted by the water motion below the ice shelf in our model calculation.  Instead, we simply model the wave/ice-shelf interaction as an oscillatory displacement at the seaward edge of the ice shelf.  

\section{Problem setup}
Figure~1 illustrates the model ice shelf we analyze.  The 2D shelf is $50$ km in length, $300$ m thick, and has basal crevasses spaced $500$~m apart.  (For clarity of display, we omitted most of the crevasses in the schematic.)  These are lengthscales typical for an ice shelf.  The left edge corresponds to the grounding line while the right edge intrudes into the ocean.  For simplicity, we assume that the displacement vanishes along the left edge and that the elastic stresses vanishes along the right edge.  The crevasses are modeled as rectangular notches into the ice shelf, with each crevasse being $150$ m in depth.  The Young's modulus of the ice shelf is $1$ GPa, the Poisson ratio is $0.3$ and the density is $1000$ kg/m$^3$. 
	\begin{figure*}	
		\centering{
		\includegraphics[width=150mm]{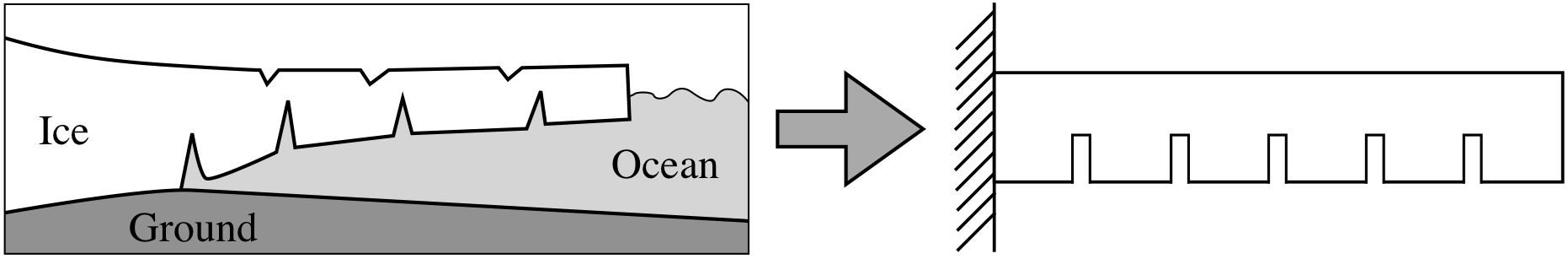}
		}
		\caption{To assess how elastic flexural waves propagate across a crevasse-ridden ice shelf (left), we analyze the normal modes of a model ice shelf (right).}
		\label{geom}
	\end{figure*}
The elastic flexural modes satisfy the equation
	\begin{equation} \label{eq:gov}
	\rho \ddot \mathbf{u} = \frac{E}{2 (1+\nu)} \bigtriangledown^2 \mathbf{u} + \frac{E}{2 (1+\nu)(1-2\nu)} \mathbf{\bigtriangledown} (\bigtriangledown \cdot \mathbf{u}),
	\end{equation}
where $\mathbf{u}(x, y, t)$ is the displacement of the ice shelf.  The left hand side of equation~(1) corresponds to the inertia of the ice shelf while the right hand side describes elastic stresses.  The first term on the right hand side is primarily associated with transverse, or bending, modes, while the second term is associated with longitudinal modes.  We use a coordinate system where $x$ is the horizontal distance along the 2D shelf and $y$ is the vertical coordinate.  To specify the normal modes completely requires boundary conditions.  Since the purpose of this calculation is to assess the leading order effect of crevasses on the mechanical response of an ice shelf, we opt to impose highly idealized boundary conditions. The left edge of the ice shelf is taken to be the grounding line and prescribed to have zero displacement.  The right edge, as well as the top and bottom surfaces of the ice shelf, experiences zero tangential and normal stresses.   (Preliminary work not included here shows that more realistic choices--say, introducing hydrostatic pressure--do not change the qualitative outcome.) 

\section{Results}
We present two types of results characterizing how the presence of a periodic array of crevasses changes the mechanical response of an ice shelf.  We first assess the effect crevasses have on the normal mode spectrum.  In this approach, instead of analyzing how a specific wave disturbance propagates along a crevasse-ridden shelf, we use the fact that the wave disturbance can be thought of as a sum of normal modes.  These correspond to standing waves, or elastic-flexural vibrations, of the ice shelf.  Our key finding is that the dispersion curve of a crevasse-ridden ice shelf is qualitatively different from that of an intact ice shelf free of crevasses.  The dispersion curve describing the wavenumber, frequency relation is not continuous, reflecting an even distribution of normal modes as a function of $f$, but instead has frequency intervals where no eigenmodes exist, i.e., a band gap.  Second, we demonstrate via a direct simulation that forcing the ice shelf at a frequency within the band gap elicits a qualitatively different response from the response when the gap is absent.  All the numerical calculations were performed using COMSOL.  

\begin{figure*}	
		\centering{
		\includegraphics[width=140mm]{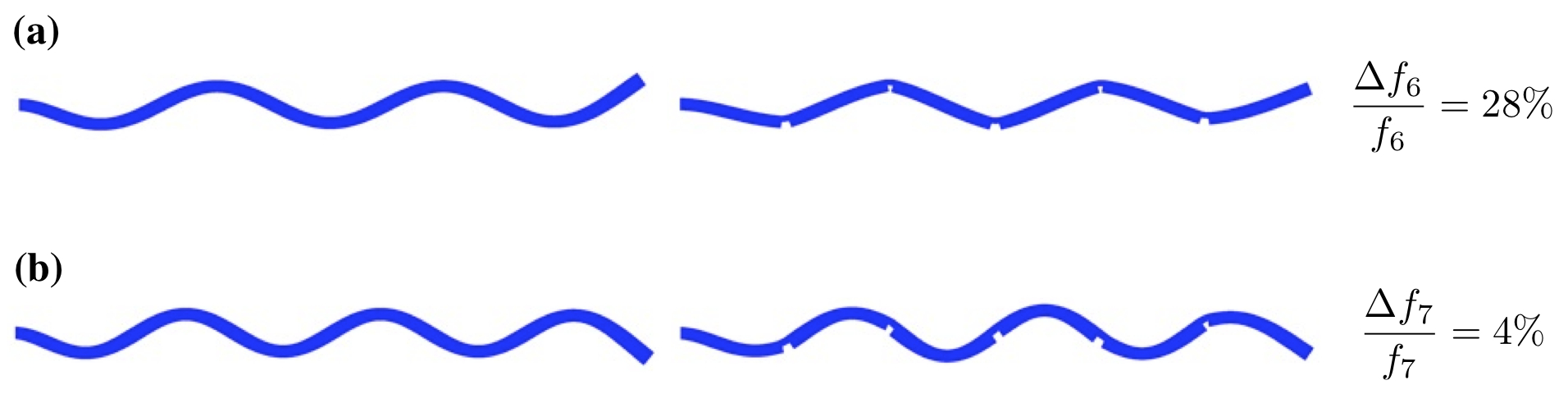}
		}
		\caption{
Conceptual demonstration calculation illustrating how the presence of crevasses in an ice shelf causes its normal mode frequencies to become unevenly distributed.   (a) The locations of the $5$ crevasses coincide with the antinodes, the positions of maximal displacement, of the $6$th normal mode. (b) In contrast, the crevasse locations coincide with the nodes, the positions of zero displacement, of the $7$th normal mode.  (In order to display the shape of the transverse modes clearly, we have vastly exaggerated the amplitudes of motion.  All the calculations assume that the displacement amplitude is much smaller than the ice shelf thickness.  The distortions in the shape of crevasses are display artifacts and not included in our normal mode calculation.) 
		}
		\label{modes}
	\end{figure*}

The impact on the normal mode distribution, being more general, is the more powerful result.  We will describe the results in two steps.  First we describe the results from a conceptual demonstration, where the ice shelf has only $5$ crevasses.  This will help build intuition for the results we obtain for the more realistic case.  In our model, the presence of the crevasses only changes the geometry of the ice shelf. It does not directly change the stress along the ice shelf boundary.  However, the crevasses do indirectly affect the stress distribution, since a section of the ice shelf with a crevasse in it bends more readily.  Intuitively, this means there exists a potential synergy between the crevasse locations and the antinodes of an elastic-flexural wave along the ice shelf.  If every antinode of the wave coincides with a crevasse, then the elastic energy of the vibration can be lowered by leaving the long stretches of the ice shelf between the crevasses nearly straight at the price of bending the small region near the crevasses more strongly.  This results in a waveform which is not sinusoidal, but instead has a ``sawtooth" shape. Figure~\ref{modes} shows that, for a $5$ crevasse ice shelf, this occurs for the $6$th normal mode.  (The transverse modes are ordered by frequency, with the lower frequencies showing up first.)   As contrast, we show also in Figure~\ref{modes} the $7$th normal mode.  This mode is essentially unaffected by the crevasses, since they lie at nodes of the wave form, where the displacement is zero.  
	
Since the presence of the crevasses essentially reduces the stiffness, or ``spring constant'', for transverse vibrations of the ice shelf, one expects that the natural vibrational frequencies of all the normal modes would decrease when the crevasses are present.  But what makes the effect nontrivial and interesting is that the frequency of adjacent eigenmodes are modified by rather different amounts. The synergy between the crevasse location and the antinode location causes the vibrational frequency of the $6$th normal mode to decrease by $28\%$.  In contrast, the vibrational frequency of the $7$th normal mode decreases only by about $4\%$.  As a result, the normal modes of an ice shelf with crevasses become unevenly distributed as a function of frequency.  

The scenario described above suggests that the crevasses are expected to affect the normal mode response most strongly when every crevasse location coincides with an antinode of the normal mode.  In other words, $\overline{\Delta f_n} = F_n - f_n / F_n$ is largest when the normal mode's wavelength is precisely double the spacing between the crevasses. In the $5$ crevasse problem, if we consider a normal mode with a shorter wavelength, say $n=20$, then only a fraction of the antinodes would coincide with the crevasse locations. As a result, the waveform is less distorted from that for an intact ice shelf. The relative frequency shift is also smaller.  The results of our simulation are consistent with this expectation.

\begin{figure}	
		\centering{
		\includegraphics[width=80mm]{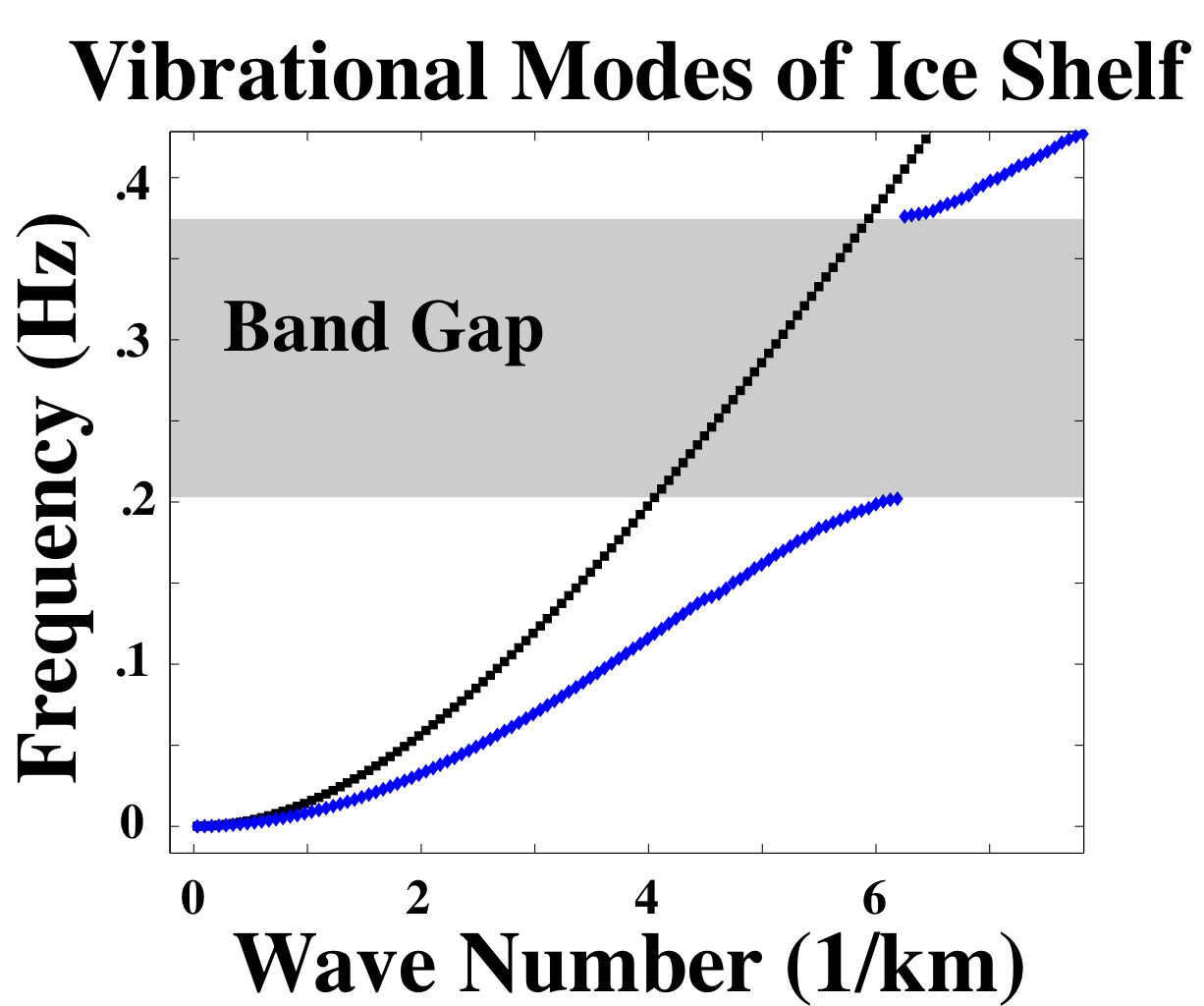}
		}
		\caption{
Dispersion curves for an intact ice shelf (black) versus our model, crevasse-ridden ice shelf.  The flexural-wave normal modes of an intact ice shelf are distributed evenly in $f$. The mode frequency increasing quadratically with the wave number $k$. For a crevasse-ridden ice shelf, the dispersion curve has its first band gap from approximately $0.2$~Hz to $0.38$~Hz.  No eigenmodes of the system exist in this frequency range. In contrast, an intact ice shelf has many eigenmodes in this frequency range. Note also that the dispersion curve for the crevasse-ridden ice shelf flattens near the edges of the band gap. This feature means that the normal modes near the band gap edge propagate significantly slower than their counterparts for an intact ice shelf. 
		}
		\label{spec}
	\end{figure}

Having examined the $5$ crevasse example, we now turn to results on the model ice shelf.  In Figure~\ref{spec} we display the dispersion curve, $f(k)$, for an intact ice shelf with the same lateral extent and thickness as our model along with the dispersion curve for the model ice shelf.  The two curves are qualitatively different.  As for the $5$ crevasse problem, the presence of the crevasses causes the normal modes to be unevenly distributed.   More importantly, we find that having a dense distribution of crevasses (having $100$ crevasses over the ice shelf instead of merely $5$) exaggerates the unevenness of the mode distribution. There is clearly a ``band gap" from $0.2$~Hz to $0.38$~Hz, corresponding to normal modes with wavelength of approximately $1$ km.  These are the vibrational modes with nearly all the antinodes located near the crevasses and are therefore most strongly affected. The crevasse-filled ice shelf has no normal modes that vibrate with a frequency within that interval.  In contrast, an intact ice shelf has many normal modes within this frequency range. Since the total number of modes has not changed significantly when we change from an intact ice shelf to a crevasse-ridden ice shelf, the normal modes that span the band gap for an intact shelves are instead concentrated into the regions near the edge of the band gap.  This is evident in the flattening of the dispersion curve $f(k)$ for the crevasse-ridden ice shelf, signifying that the edge regions supports an unusually large number of normal modes with nearly identical natural vibration frequencies. 

In the limit of a shelf whose lateral extent is much larger than the spacing between the crevasses, the group velocities of these eigenmodes approach zero since $f(k)$ is nearly horizontal near the edge. As a result, energy from the incident wave does not propagate into the ice shelf but is instead reflected out into the ocean.  This behavior is, mathematically, completely equivalent to the coherent back scattering scenario described for a propagating wave.  The only difference is that in the normal mode analysis it is natural to consider how crevasses interact with a standing wave while the coherent back scattering scenario is the natural description for a propagating wave.

\begin{figure}	
		\centering{
		\includegraphics[width=80mm]{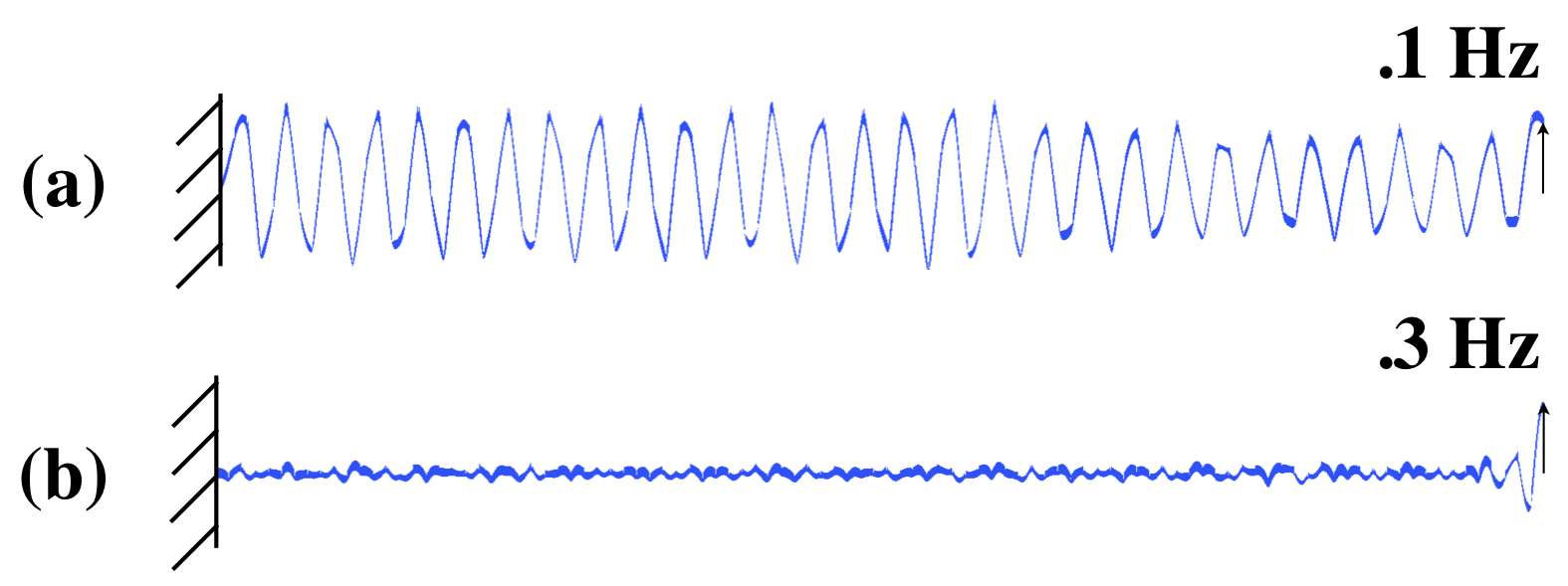}
		}
		\caption{ 
Response of a crevasse-ridden ice shelf to prescribed oscillatory displacement of its seaward edge. We have exaggerated the amplitudes of the vertical displacement by the same scale factor in both figures in order to display the ice shelf response clearly.  (a) Response to driving frequency which lies outside the band gap:  when the ice shelf edge is oscillated at $0.1$~Hz, which lies outside the band gap, the response resembles that of an intact ice shelf.  The forcing excites several normal modes. The entire ice shelf to vibrates with more or less the same amplitude. (b) Response to driving frequency which lies inside the band gap:  when the ice shelf edge is displaced at $0.3$~Hz, which lies {\it within} the band gap, the forcing excites normal modes from the edges of the band gap.  The amplitude of the ice shelf motion attenuates strongly as we move away from the edge. 
		}
		\label{driv}
	\end{figure}

Finally we assess how the band gap changes the response of an ice shelf to disturbances from ocean waves.  We mimic the ocean wave disturbance as a prescribed oscillatory displacement 
\begin{eqnarray}
	u_x = 0; & & u_y  = \sin(2\pi f t)
\end{eqnarray}
at the seaward edge of the ice shelf.  Figure~\ref{driv} contrasts the response for $f=0.1$~Hz, a frequency outside the band gap and $f=0.3$~Hz, a frequency inside.  The time of the simulation snapshot are taken at (a) $t=303$~s and (b) $t=297.7$~s.  These times are chosen so that the unperturbed waves in both simulations are given enough time to traverse the length of the ice shelf and return significantly.  Slightly different times are taken so that the driving displacements have the same phase.  When the disturbance frequency $f$ lies outside the band gap (Figure \ref{driv}a), the response is qualitatively the same as that of an intact ice shelf.  The edge disturbance excites elastic-flexural waves whose frequencies are close to the driving frequency. These waves initially travel from the seaward edge towards the grounding line, are reflected at the grounding line, travel back outwards towards the seaward edge. As a result, the whole ice shelf is set into oscillatory motion. 

Inside the band gap (Figure~\ref{driv}b), the response is very different.  Since no elastic-flexural wave normal modes exist with natural frequencies near the driving frequencies, the response of the ice shelf is dominated by normal modes from the edges of the band gap.  As a result, disturbances introduced at the ice shelf edge propagate inwards very slowly.  Over a longer time-scale, the response of the ice shelf continues to be qualitatively different. The amplitude of the ice shelf motion decays strongly as one moves away from the edge.  Most of the ice shelf interior is undisturbed.  This striking attenuation, reminiscent of an evanescent wave, can be understood, at least qualitatively, as a consequence of the coherent back scattering effect described in the introduction. Each time an elastic-flexural wave hits a crevasse, it is partially transmitted and partially reflected.  When the wavelength is mismatched from the spacing between the crevasses, the reflected waves from the different crevasses arrive at the seaward edge of the ice shelf with different phases, thereby interfering with each other destructively.  When the wave length of the normal mode is roughly twice the spacing between the crevasses, however, the reflected waves from the different crevasses can arrive at the seaward edge of the ice shelf in phase, thereby interfering with each other constructively.  As a result, almost all the energy of the incident wave is reflected back into the ocean, instead of being dispersed through the ice shelf.  
In the limit that the ice shelf extent becomes infinitely large relative to the spacing between the crevasses, the reflections from all the crevasses add constructively to create a true evanescent wave, one that decays exponentially with distance from the seaward edge.  In reality, the number of crevasses on an ice shelf is large but finite, so the dynamical response only approximates an evanescent wave. Forcing the ice shelf at a frequency which lies within the band gap excites a response from the ice shelf with the largest amplitudes from frequencies at the edges of the band gap.

If the ice shelf is exposed to a disturbance whose power is distributed evenly across a range frequencies that span the band gap, then the response of the ice shelf will then have distinctive peaks in the power spectrum, with each peak corresponding to one edge of the band gap.
 
\section{Discussion} 
 
This work is a first attempt to assess quantitatively how the mechanical response of an ice shelf is altered by the presence of crevasses.  To keep the analysis tractable, we chose to idealize the crevasses, which in reality are disordered, uneven in depth and variable in material properties, as spatially periodic and identical.  We have checked that introducing a small amount of disorder, such as created by varying average spacing, does not change any of the main features reported here.  We have also checked that introducing a gentle taper in the ice shelf thickness, or altering the shape of a crevasse from a rectangular notch to a wedge, or including the hydrostatic pressure term, do not change our results. In short, the band gap is a robust feature and emerges whenever an ice shelf has a regular array of crevasses. 
  
We see several future directions that naturally extend this work.  First, we have included the wave-shelf interaction in the most simple manner possible, as an oscillatory displacement of the ice shelf edge.  It would be interesting to model the interaction more realistically by solving for the motion of the water under the ice shelf self-consistently with the ice shelf motion.  In this second scenario, the mechanical disturbance is not confined to the ice shelf edge, but instead is distributed throughout the interior.  This calculation may give us insight into how a crevasse-ridden ice shelf responds to tidal motion. Coupling the ocean and ice shelf interaction will also give us a more precise indication of when the crossover occurs between a response dominated by water motion and one dominated by the inertia of the crevassed ice shelf.

Second, we emphasize that the emergence of the band gap in the dispersion curve depends fundamentally on a geometrical interplay between the wavelength of the elastic-flexural normal mode and the distribution of crevasses.  Thus, disorder and variability, if sufficiently strong, will introduce qualitative changes.  Works to characterize this second regime of dynamical response are underway. 

Finally, the fact that the same ice shelf can respond in a dramatically different way depending on whether it experiences an edge disturbance inside or outside the band gap may be relevant for ice shelf collapse. To see why, let us suppose that the crevasse distribution on an ice shelf can evolve over a relatively long time-scale so that, typically, the environmental disturbances perturb the ice shelf only within its band gap range.  If this assumed evolution exists, then a sudden thinning of the ice shelf may be catastrophic if the reduction in the ice shelf thickness manages to shift the band gap appreciably, since the frequencies of the normal modes vary strongly with the ice shelf thickness. Since forcing from the ocean waves are unchanged, a shift in the frequency interval for the band gap would switch the response of the ice shelf from the within-band gap response, in which most of the ice shelf interior is shielded from wave disturbances, to the out-of-bang gap response, where the entire ice shelf experiences the effect of the waves. Testing this idea requires field work that more fully characterizes the elastic-flexural wave response of ice shelves, as well as an understanding of whether, and how, crevasses evolve on an ice shelf. 

\section{Conclusion}

The propagation of elastic-flexural waves through an ice shelf can be strongly modified by the presence of crevasses.  We demonstrate this feature by calculating the normal modes of an ice shelf under the assumption that only inertia of the ice shelf and elastic stresses are significant.  Unlike the case for an intact ice shelf, the normal modes for a crevasse-ridden ice shelf are unevenly distributed as a function of frequency.  For a model ice shelf with dimensions and crevasse spacings consistent with field observations, there exists a ``band gap" from $0.2$ to $0.38$ Hz, an interval of frequency where no normal modes are possible.  This behavior contrasts sharply with a crevasse-free ice shelf, which supports many normal modes within that frequency range.  This band gap structure gives rise to a qualitatively different mechanical response. Driving a crevasse-ridden ice shelf at its seaward edge with a frequency in the band gap causes the disturbance to be confined near the edge of the ice shelf, reminiscent of an evanescent (non-propagating) wave.  Most of the interior of the ice shelf is left quiescent.  In contrast, a crevasse-free ice shelf shows a response that is evenly distributed throughout the interior.    More generally, the existence of a band gap implies that the response of a crevassed ice shelf exposed to wave forcings that cover a broad frequency range should show peaks in the power spectrum, with each peak corresponding to one edge of the band gap.
\\ 

\noindent{\bf ACKNOWLEDGEMENTS}  

We thank Dorian S. Abbot, Justin C. Burton, L. Mac Cathles IV, Nicholas Guttenberg and Sidney R. Nagel for encouragement and insightful feedback. This work is supported by the US National Science
Foundation under grants ANT-0944193, OPP-0838811 and CMG-0934534. J. Freed-Brown was supported by the Materials Research Science and Engineering Center (MRSEC) at the University of Chicago (DMR-0820054).

\end{document}